\documentstyle[twocolumn,epsfig]{mn2e}
\newcommand{\be}{\begin{equation}}
\newcommand{\ee}{\end{equation}}
\newcommand{\ba}{\begin{eqnarray}}
\newcommand{\ea}{\end{eqnarray}}
\def\la{\lower0.6ex\vbox{\hbox{$ \buildrel{\textstyle <}\over{\sim}\ $}}}   
\def\ga{\lower0.6ex\vbox{\hbox{$ \buildrel{\textstyle >}\over{\sim}\ $}}} 
\def\Msun{{\rm M}_{\odot}}

\begin{document}

\title[Peculiar velocity power spectrum]{A new method of measuring the cluster peculiar velocity power spectrum}

\author[Zhang, Feldman,  Juszkiewicz \& Stebbins]{Pengjie Zhang$^{1,2,\star}$, Hume
A. Feldman$^{3,\dagger}$,  Roman Juszkiewicz$^{4,5,\ddagger}$ \& Albert Stebbins$^{2,*}$ \\
$^1$Shanghai Observatory, Chinese Academy of Science, Shanghai, 200030 China\\
$^2$NASA/Fermilab Astrophysics Center, Fermi National Accelerator Laboratory, Batavia, IL 60510--0500\\
$^3$Dept. of Physics \& Astronomy, Univ. of Kansas, Lawrence, KS 66045, USA\\
$^4$Institute of Astronomy, Zielona G{\'o}ra University, 65--516 Zielona G{\'o}ra, Poland\\
$^5$Copernicus Astronomical Center, 00--716 Warsaw, Poland\\
emails:$^\star$pjzhang@shao.ac.cn; $^\dagger$feldman@ku.edu;$^\ddagger$roman@camk.edu.pl;$^*$stebbins@fnal.gov}

\maketitle

\begin{abstract}
We propose to use spatial correlations of the kinetic
Sunyaev--Zeldovich (KSZ) flux as an estimator of the peculiar velocity
power spectrum.  In contrast with conventional techniques, our new
method does not require measurements of the thermal SZ signal or the
X--ray temperature. Moreover, this method has the special advantage  that
the expected systematic errors are always sub--dominant to statistical
errors on all scales and redshifts of interest. We show that future
large sky coverage KSZ surveys may allow a peculiar velocity power
spectrum estimates of an accuracy reaching $\sim 10\%$. 
\end{abstract}
\noindent{\it Subject headings}: cosmology: large scale structure; theory--cosmic microwave background; cosmology: observations; cosmology: theory; dark matter; distance scale

\section{Introduction}
\label{sec:intro}

Peculiar velocity--distance galaxy surveys have provided us with maps
of line--of--sight peculiar velocity fields,
$v({\bf x}) = {\bf V}\cdot {\bf x}/x$, where ${\bf V}$
is the peculiar velocity at position ${\bf x}$. Such surveys were
successfully used in the past to constrain 
power spectra of initial density fluctuations and to
estimate the cosmological density parameter, $\Omega_m$ 
\cite{Vittorio,Groth,daCosta,Feldman03a,Feldman03b,sarkar07,wf07,fw08}. For an
overview of the subject, see \cite{Strauss} and \cite{Courteau}.
However, all these velocity--distance surveys 
are limited by errors in the range of 15 to 20\% of the estimated
distance. Measurements based on type Ia supernovae have slightly lower error
per source, due to smaller dispersions in type Ia supernovae intrinsic
luminosities (see, e.g. \cite{Hui06,Gordon07} and references therein for theory
aspects and  \cite{Bonvin06,Haugboelle06,Neill07,Wang07,wf07} and
references   therein for recent measurements).
In contrast, for cluster line--of--sight 
peculiar velocities, $v$, derived from the kinetic Sunyaev
Zeldovich (KSZ) effect, the errors grow much less rapidly
with distance. KSZ cluster surveys
may open new possibilities for studying large--scale flows
\cite{Haehnelt96,Kashlinsky00,Aghanim01,Atrio--Barandela04,Holder04}
and cosmology \cite{Bhattacharya07,Fosalba07}.
The currently established method of recovering $v$
from the SZ data requires extra measurements of the
cluster temperature and its Thomson optical depth, $\tau$.
Other complications include the relativistic thermal Sunyaev Zeldovich
(TSZ) effect and  
the KSZ signal generated by internal motions, which is added to the signal
from cluster's bulk velocity, resulting in a limit in accuracy
in determining $v$, of the order of $\sim 200$
km/s \cite{Knox03,Aghanim04,Diaferio04}. 

In this paper we propose to estimate the spatial 
two--point correlation function of the KSZ flux from the 
data and then infer the velocity power spectrum. 
Contaminations to the cluster KSZ flux have various clustering
properties that can be applied to disentangle the 
signal from its contaminants (see also \cite{zhang04}). We will show that
at $z\ga 0.3$, the systematics become 
sub--dominant and the statistical errors, at $\sim 10\%$ level for
South Pole telescope (SPT\footnote{http://spt.uchicago.edu/, specifically. see the SPT White Paper at this site.}) \cite{SPT04},
dominate.
Throughout this {\it letter}, we adopt the cosmology with
$\Omega_m=0.3$, $\Omega_{\Lambda}=1-\Omega_m$ and  
a normalization parameter $\sigma_8=0.9$.

\section{The flux power spectrum}
\label{sec:FPS} 

The KSZ cluster surveys directly measure the sum of the cluster KSZ flux
$S_{\rm KSZ}$ and
various contaminants, such as intracluster gas internal
flow,  radio and IR point sources, primary 
cosmic microwave background (CMB),
cosmic infrared background (CIB), etc. 
The signal is \cite{Sunyaev80} 
\be
S_{\rm KSZ}
\; = \;\frac{\partial B_{\nu}(T)}{\partial T} \Delta T_{\rm KSZ}\;,
\ee
where $B_{\nu}(T)$ is the Planck function. At $\nu\sim 217 $Ghz,
$\partial B/\partial T=540 \ {\rm Jy}\ {\rm sr}^{-1}\ \mu K^{-1}$. 
The KSZ temperature  fluctuation can be expressed as
\be
\label{eqn:dTKSZ}
\Delta T_{\rm KSZ} = 2.7{\rm K}\, \langle \tau \rangle \,(v/c)\, = 
9 \mu K \,  v_{100}\,\langle\tau\rangle /0.01 ,
\ee
where $\langle\tau\rangle$ is the optical depth averaged over the solid 
angle $\omega$, while $v_{100}\equiv v/(100 {\rm km}/s)$. We will also use
$S_{100}\equiv S_{\rm KSZ}(v)/S_{\rm KSZ}(100\,{\rm km}/s)$. These quantities are,
respectively, the ``normalized'' peculiar velocity and KSZ flux, scaled to  
$v = 100\,{\rm km}/s)$.  
The value $\, \langle\tau\rangle = 0.01$ is typical
for cluster models considered in numerical simulations (see e.g. \cite{antonaldo}
and references therein).  
Since at
$\nu\sim 217$ Ghz, the non--relativistic part of the TSZ effect, which is one of
the major contaminants of the KSZ,  vanishes \cite{Zeldovich69}, we focus on this
frequency.  Optical follow--up of KSZ surveys such as dark energy 
survey\footnote{http://www.darkenergysurvey.org/} 
will measure cluster redshift $z$ with uncertainty  $\la
0.005$.\footnote{The photo--$z$ of each galaxy has dispersion $\sim 0.05$. 
Clusters have $\ga 100$ galaxies and thus the determined $z$
dispersion is $\la 0.005$.} The $z$ information allows the
measurement of the 3D auto correlation of measured cluster KSZ flux
\be
\label{eqn:xis}
\xi_{S}(r) \equiv \langle \Delta S_i \Delta S_j \rangle
= \langle S_{{\rm KSZ},i}S_{{\rm KSZ},j} \rangle+\cdots\ , 
\ee
where $\Delta S_i =  S_i-\overline{I}\omega_i\;$,
$S_i$ and $\omega_i$ are the total flux and
the solid angle subtended by the 
$i$--th cluster; $r = |{\bf x}_i - {\bf x}_j|$ is the comoving
separation of a pair of clusters, and $\bar{I}=\sum S_i/\sum \omega_i$ is the mean flux 
intensity. The total flux is a sum of the KSZ signal, $S_{{\rm KSZ},i}$, and various
noise terms, which we study below.  

The correlation $\langle S_{{\rm KSZ},i}S_{{\rm KSZ},j} \rangle$ is
cluster number density weighted and thus
\be
\langle S_{{\rm KSZ},i}S_{{\rm KSZ},j}
\rangle\propto \langle (1+\delta_{C}(M_i))v_i\; (1+\delta_{C}(M_j))v_j \rangle\ \; ,
\ee
where $\delta_{C}(M)=b_n(M)\delta\,$ is the cluster number overdensity, while
$\delta$ is the matter overdensity. Here we use a simple biasing model,
where the parameter $b_n$ is determined by the cluster mass $M$
\cite{Mo96} (see also refs. \cite{K87,K91}). 
The quantity above can be written as 
\be
\langle (1+\delta_i)v_i\; (1+\delta_j)v_j \rangle\ \; = \; 
\langle v_i v_j\rangle + \langle \delta_i \delta_j v_i v_j\rangle \; ,
\ee
where the third order moments cancel because of statistical isotropy and homogeneity 
which reflects the symmetry expected in the large scale pattern of the cluster distribution:
$\, \langle v_i v_j \delta_i\rangle = - \langle v_i v_j \delta_j\rangle$. 
The fourth moment can always be written as a sum of products of second
moments of all possible pairs plus the irreducible or connected fourth moment:
$\langle 1234\rangle = \langle 12 \rangle \langle 34 \rangle + \langle 13 \rangle
\langle 24 \rangle  + \langle 14 \rangle \langle 23 \rangle + \langle 1234 \rangle_c $.

The signal we are interested in, 
$\langle S_{{\rm  KSZ},i}S_{{\rm KSZ},j} \rangle$, is directly related to the
peculiar velocity 
logarithmic power spectrum $\Delta^2_{v}$. Here and below,
$\Delta_Q^2 \equiv P_Q(k,z)k^3/(2\pi^2)$, where $k$ is the
comoving wavenumber, $z$ is the redshift and $P_Q$ is the proper power
spectrum of the random field $Q$. 
The variance is $\langle Q^2\rangle = \int_0^\infty dk\,\Delta^2_Q/k$. 

The nonlinear scale at 
$z\sim 1$ is $k\simeq0.5h/$Mpc, hence for wavenumbers
$k< 1h/$Mpc we can apply the weakly nonlinear perturbation theory. 
We also assume here that  
to first order, the density and velocity are Gaussian random
fields, related by the expression 
\be
\label{eqn:v}
\Delta_v (k,z)\;= \;f\,H\,a\, \Delta_m/\,H_0\,k\,\sqrt{3}\, ,
\ee
where $a$ is the scale factor, $f\equiv d\ln D/d\ln a$, $H = H(z)$ is the Hubble parameter
at redshift $z=1/a-1$ and $H_0 = H(0)$, and $D(z)$ is the usual linear growth factor for
density perturbations. $\Delta^2_m$ is the logarithmic power spectrum of
the dark matter density fluctuations derived using
the standard transfer function \cite{Bardeen86} and assuming $\Omega_m = 0.3$,
$\Omega_{\Lambda} = 0.7$, $\sigma_8 = 0.9$ and $H_0 = 70$ km/s/Mpc.
Under these assumptions, to lowest order, the connected fourth moment vanishes
because of Gaussianity while the nonlinear, mode coupling contribution to the signal 
is well approximated by the expression \cite{Jim}
\be
\label{eqn:dvdv}
\Delta^2_{\delta  v}({\bf
k})\; \approx \;
\Delta^2_{m}(k,z)\, \sigma^2_v(z) \; /\; 2 \;, 
\ee
where $\sigma_{v}^2 = \int_0^{\infty}dk\,\Delta_v^2 /k$ is the
line-of-sight velocity  dispersion. 
We also assume that the two lines of sight are almost parallel.
The accuracy of this assumption is
better than $\sim 1\%$ for angular separations
$< 10^{\circ}$.

Bulk flows are induced by the large
scale gravitational potential, so the velocity field
decouples from the small scale density field and 
the last expression in Eq.(\ref{eqn:dvdv}) remains valid
and agrees with N-body simulations
in the nonlinear regime, at redshifts $Z < 1$ and wavenumbers
$k> 1h/$Mpc if the linear $\Delta^2_m$ is replaced by
its nonlinear analog \cite{Jim}.

To express these results in terms of the cluster number-density
fluctuations $\delta_C (M)$, we will again turn to a simple
Press-Schechter prescription, described earlier.  
The KSZ signal is then 
\be
\label{eqn:KSZ}
\Delta^2_{\rm KSZ}(k) = s_0^2 \left(\Delta^2_{v}(k)+
{\overline{b_n}}^2\Delta^2_{v \delta}\right)\;,
\ee
where
\ba
s_{N}\equiv\int_{M_*}^\infty b_n^{N}(M)S(M)(dn/dM)dM,\,N=0,1\, ,
\nonumber
\ea 
and $\overline{b_n}\equiv s_1/s_0$. The quantities $\Delta_v^2$ and 
$\Delta^2_{v \delta}$ are the logarithmic power spectra of spatial
correlation functions $\langle v_1v_2\rangle$ and 
$\langle \delta_1v_1v_2\delta_2\rangle$, introduced above.
To describe the density field, we use
the cluster mass function, written as 
$dn/dM$, and $M_*$ is the
lower mass threshold of SZ selected clusters. 
For SPT, $M_*\sim 10^{14} \Msun$.
The redshift dependence of
$M_*$ \cite{Holder00} is weak and we assume a constant $M_*$.
There is no biasing assumption for velocities
because we assume that the galaxies trace
the dark matter velocity field,
in agreement with observations \cite{Feldman03a}. 
To calculate the effective
cluster bias, we follow the halo model of
\cite{Mo96}:
$b_n(M)=1+(\nu^2-1)/(\delta_c/D)$, where $\delta_c\simeq 1.686$,
$\nu=\delta_c/(D\sigma(M))$ and $D(z)$ is the linear density growth
factor.  The quantity $\sigma(M)$ is
the rms density fluctuation in a spherical volume containing mass $M$
and we have  $\sigma(M_*)\simeq \sigma_8=0.9$. For the present
estimate we use the approximation 
$\overline{b_n} =  b_n(M_*)$.  At present time 
$\overline{b_n}(M_*, 0) \approx 3$.
This value is consistent with most recent measurements of the correlation function
of optically selected galaxy clusters in the Sloan Digital Sky Survey \cite{Josh}.
This is, indeed, a simplified version of the biasing scheme, however for the scales of interest in this paper it is justified.

\begin{figure}
\epsfxsize=9cm
\epsffile{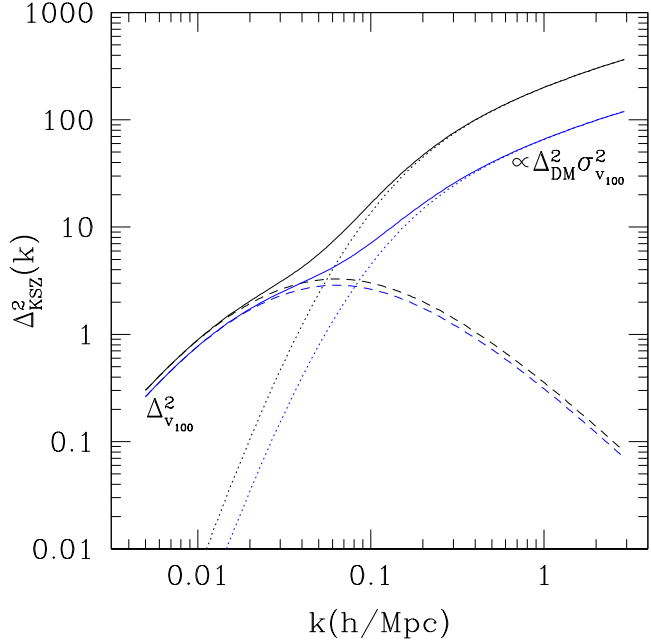}
\caption{The normalized KSZ cluster flux power spectra at $z=0$ (upper
  black solid
line) and $z=1$ (lower blue solid line) respectively. The power spectrum,
$\Delta_{\rm KSZ}^2(k)$ is expressed in dimensionless units, defined
in \S 2. The dashed and dotted lines show the
contributions from the $vv$ and $v\delta v\delta$ terms, respectively. 
The solid line shows the sum of these two terms. 
$\sigma_{v_{100}}^2 \equiv \sigma_v^2/(100\,{\rm km}/s)^2$. The normalization is $\Delta^2_{KSZ}/s_0^2$.
\label{fig:v}}
\end{figure}

In Fig. \ref{fig:v} we show the normalized KSZ 
flux power spectrum at $z=0$ and at $z=1$.
It is clear that the $vv$ contribution  
dominates at small $k$ whereas the $\delta v\delta v$ term dominates in the
high $k$ tail. 
$\Delta^2_{v}$ peaks at $k\sim 0.05h/$Mpc, because of the $k^2$ denominator, while 
$\Delta^2_m$, associated with contaminants traced by point sources, keeps 
decreasing toward large scales. On the other hand, 
$\Delta^2_{\rm KSZ}$ increases with $k$ 
because of the presence of the mode--coupling term.
Note also, that compared to
$\Delta^2_m$, $\,\Delta^2_{\rm KSZ}$ has a weak
$z$ dependence at $k\la 0.03h/$Mpc.

We have not considered the effect of redshift distortion. 
It reduces the power of  the $\Delta^2_v$ part by a
factor of $\sim \exp[-k^2\sigma_v^2/H^2(a)]$. Since $\sigma_v\sim 200$ km/s, the
redshift distortion can be neglected  at $k\la 0.2 h/$Mpc. On the
other hand, the redshift distortion increases the power of the
$\Delta^2_{v\delta}$ part in the linear regime \cite{Kaiser87}. 
Since $\Delta^2_{v\delta}$ 
dominates over $\Delta^2_v$ at $k\ga 0.1h$/Mpc (see Fig. \ref{fig:v}), neglecting
the redshift distortion tends to underestimate the signal $\Delta^2_{\rm KSZ}$
and thus overestimates the relative errors. 

\section{Noise sources}
\label{sec:error}

To study the detectability of  $\Delta^2_{\rm KSZ}$,
let us consider the South Pole telescope experiment$^6$, planned for the near future.
The SPT will cover the $217$ Ghz range and about
$4000\  {\rm deg}^2$ with arc--minute resolution. 
It is expected to observe more than 20,000 clusters with masses greater than $10^{14} \Msun$
for the thermal SZ signal. Virtually all the TSZ clusters have flux measurements at 220 Ghz band, a fraction of this flux comes from KSZ effects. 
The corresponding KSZ flux may be too weak to detect cluster by cluster, however, our method provides a way to extract this KSZ signal at 220 Ghz band 
by statistically weighting all fluxes.
Thus, cluster finding is done by combining all bands (90, 150, 220 and 270 Ghz).Ê Given these TSZ selected clusters, we can measure 
the CMB flux enclosed in the same region at 220 Ghz. The measured flux is the sum of KSZ, residualÊ TSZ, CMB and various other contaminations. 
Indeed, for many individual clusters, the KSZ signal is overwhelmed by other contaminants, however, the correlation approach described here is able to 
separate the KSZ signal from various sources of noise. 

In the following we discuss the three
dominant error sources in the following subsections.
We assume redshift bin size $\Delta z=0.2$ and $k$
bin size $\Delta k=0.4k$. The statistical errors scale as $\Delta
k^{-1/2}$.

\subsection{Diffuse foregrounds and backgrounds}
\label{subsec:fb}
Clusters have typical bulk velocity around several hundred kilometer
per second, so the typical KSZ signal is $\Delta T_{\rm KSZ}
\sim 20 \mu K \langle\tau\rangle /0.01$.  Primary CMB, which is
indistinguishable from the KSZ  
signal in frequency space,  has intrinsic temperature
fluctuation $\Delta T \sim 100 \mu$K. The cosmic infrared background
(CIB) has mean  $T\sim 20 \mu$K, if scaled to $217$ Ghz 
\cite{Fixsen98}.  A cluster KSZ filter 
can be applied to  remove the mean backgrounds while keeping the KSZ
signal. This filter must strongly match the cluster
KSZ profile while having  zero integrated area. Since the angular sizes
of clusters are 
usually close to 
several arc--minutes, such a 
filter 
peaks at multipole $l$ around several thousands and thus 
removes the dominant CMB signal, concentrated near
$l\la 1500$.  For such filters, at $\sim 217$ Ghz, the galactic
synchrotron, dust emission,  free--free foregrounds, the radio and TSZ 
backgrounds are all negligible due to their frequency or scale dependence
(see, e.g. \cite{Wright98, Bennett03}). Thus we only discuss the 
contaminations of  the primary CMB, CIB and background  KSZ. 

The optimal filter can be constructed from the intracluster gas
profile  inferred from the TSZ survey.  Without loss of generality, 
we choose the electron density profile as 
$n_e(r)\propto (1+r^2/r^2_c)^{-1}$ and a matched filter 
$W_{\ell} = 6(\ell/\ell_f)^2\exp(-(\ell/\ell_f)^2)$. 
For these particular choices, the filtered KSZ temperature, 
$\tilde{\Delta} T_{\rm KSZ}$, peaks at $\ell_f \approx 1.1/\theta_c
\approx 3800
(1^{'}/\theta_c)$, where $\theta_c$ is the angular core radius, and
the peak value is $\tilde{\Delta} T_{\rm KSZ}\simeq \Delta T_{\rm KSZ}\simeq 
9 v_{100} [\langle\tau\rangle/0.01]\mu K $. We
adopt this $\ell_f$ and $\theta_c=0.4 h^{-1}{\rm Mpc}/\chi(z)$ to estimate
the noise. Here, $\chi(z)$ is the comoving angular distance.   

The correlations of filtered
backgrounds (with zero mean flux), originating from both their intrinsic
and cluster correlations, are 
\be
\tilde{\xi}_{b}(r)\approx \left[\textstyle{\int} \omega(M)
dn(M)\right]^2\left[1+\overline{b_{n}}^2 \xi_{m}(r)\right]
\tilde{w}_{b}(\theta)\ ,
\ee
where $\theta = r/\chi$, while $\tilde{w}_{b}(\theta)$ is the
corresponding (filtered) background
angular correlation function and 
$\omega(M)$ is the solid
angle subtended by a
cluster of mass $M$ at redshift $z$. Finally,
$\xi_{m}$ is the
dark matter correlation function.

The same cluster survey measures both $\int  
\omega dn$ and $\overline{b_n}^2\xi_{m}$, while the CMB
$C_{\ell}$'s are known from observations \cite{Hinshaw03}.
Alternatively, for the cosmological model assumed here, 
the expected $C_{\ell}$'s can be also
calculated from first principles \cite{CMBFAST}. Thus the CMB
contamination can be easily subtracted from the correlation
estimator, leaving the statistical errors from
intrinsic CMB fluctuations over cluster regions as the only CMB contaminant. 

\begin{figure}
\epsfxsize=9cm
\epsffile{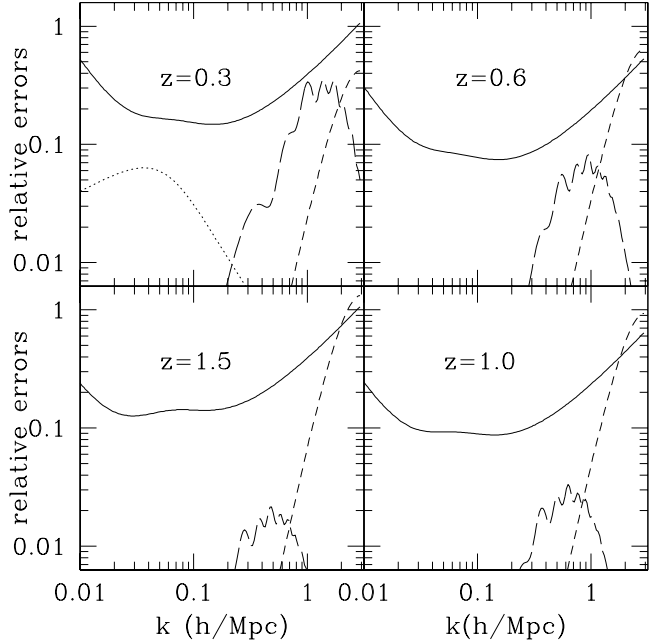}
\caption{The dominant errors of the cluster KSZ
power spectrum measurement. The solid lines are the errors caused by
the cosmic variance and shot noise. The short dash lines are the
systematic errors caused by the CIB and KSZ background. The long dash
lines are the statistical errors caused by the primary CMB. The dot
lines (which are  visible only in the upper left panel) are the
statistical errors caused by sources associated clusters. Systematic
errors are sub--dominant at $k\la 1h/$Mpc.   
\label{fig:error}}
\end{figure}

In principle, the CIB and KSZ contaminations can be removed as well. But
since both the amplitude and shape of CIB and KSZ power spectra are highly
uncertain, we do not attempt to subtract their contribution from the
correlation estimator. The CIB power spectrum is $C_{\ell}^{(1)}
\ell^2/(2\pi)\simeq (4 \mu {\rm K})^2 (\ell/10^3)^{0.7}$ 
and the  KSZ power spectrum is 
$C_{\ell}^{(2)}\ell^2/(2\pi)\simeq (2.7 \mu {\rm K})^2$ \cite{knox01}. The
resulting upper limit on the fractional systematic error is  
\be
\eta(k) \;\approx \;\frac{(1+A_k)
\sum C^{(J)}_{\ell}W_{\ell}^2\ell^2}{
2\pi\,B_k\,[9  
(\langle\tau\rangle/0.01)\mu K]^2}\;,
\ee
where $J=1,2$ is the summation index; the spherical harmonic number, $\ell= k\chi$,
is fixed, $A_k\equiv \overline{b_{n}}^2 \Delta^2_m$, while $B_k\equiv
\Delta^2_{v_{100}}+\overline{b_{n}}^2\Delta^2_{v\delta 100}$.
Where $\Delta^2_{v\delta 100}$
is the $\Delta^2_{v\delta}$ term, expressed
in units of $(100\, {\rm km}/s)^4$. Clearly, $\eta(k)$ 
is at most several percent at $k\la 1h/$Mpc (see Fig. \ref{fig:error}).

\subsection{Sources associated with clusters}
\label{subsec:clst}

The map filter does not remove the contaminations
from sources associated with clusters. The flux of cluster radio 
and IR sources is $\sim 10^3{\rm Jy}/{\rm sr}$ at $217$ Ghz \cite{Aghanim04}.  
Using multi--frequency information and  resolved source subtraction, one is
likely able to subtract much of these contaminations. Rather
conservatively, we assume that, at $\nu\sim 217 $Ghz, the total 
flux contributed by the sources associated with 
clusters is less than $\sim 5\times 
10^3 {\rm Jy}/{\rm sr}$ at $z=0.5$ and scale it to other redshifts
assuming no intrinsic luminosity evolution.  At $217$Ghz, the
non--relativistic TSZ vanishes. The relativistic
correction of cluster TSZ effect shifts the cross over point slightly to higher
frequency and thus in principle introduces a residual TSZ
signal in $\sim 217$Ghz band. But the correction to $\Delta T$ is
generally only several percent of the TSZ at Rayleigh--Jeans regime 
\cite{Stebbins97, Itoh98}. 

The mean flux of the cluster sources, $S_{\rm cl}$, is subtracted in
our estimator 
(Eq. \ref{eqn:xis}). Since the cluster thermal energy, IR and radio
flux  should be  mainly  determined by local processes,  one can omit
the possible large--scale  correlations of these  quantities. Thus these
sources do not cause systematic errors. But since 
the mean square of the flux fluctuations, caused by clusters, 
$\langle \delta S_{\rm cl}^2\rangle$, does not vanish, 
they do contribute a shot noise, which will be estimated in \S
\ref{subsec:stat}.  Further, they contribute a statistical noise term
proportional to the cluster clustering, whose expected fractional error is $\eta(k)\approx 2\pi\,A_k\langle \delta S_{\rm cl}^2\rangle
/S^2_{100}B_k\,\sqrt{Vk^2\Delta k}\;$, for a survey of volume $V$. This can be rewritten as
\be
\eta(k)\; \approx \; 5\times 10^{-3}\frac{\langle \delta S_{\rm cl}
^2\rangle}{zS^2_{100}}\sqrt{\frac{\Delta z}{0.2} \frac{\Delta k/k}{0.4}} \; . 
\ee
It is reasonable to
assume that  $\langle \delta S^2_{\rm cl}\rangle^{1/2}\la \langle
S_{\rm cl}\rangle$. Thus  the error caused by
the sources  associated with clusters  is negligible at  all
scales and redshifts (See Fig. \ref{fig:error}).

\subsection{Cosmic variance and shot noise}
\label{subsec:stat}
The signal intrinsic cosmic variance dominates at large
scales. The number of clusters is limited, 
so the shot noise is large, even in the linear regime. 
We believe that there are five dominant sources of shot noise.
The shot noise power spectrum is  then given by 
\be
\tilde{\Delta}^2_{\rm shot}\;=\;\frac{k^3}{2\bar{n}
\,\pi^2}\;\sum_{J=1}^5 \;\tilde{\sigma}^2_{J}\;, 
\ee
where the summation index $J$ corresponds to mean square noise
values of discrete sources, associated with clusters $(J=1)$,   
the instrumental noise, the residual
flux fluctuations of primary CMB, CIB, and the background KSZ
($J=2$ to 5, respectively).
$\bar{n}$ is the mean  number density of observed clusters, which can
be calculated given the halo mass function, the survey 
specification and the gas model. We use
the expression $\bar{n}(z)=3\times 
10^{-5}/(1+z)^3 (h/{\rm Mpc})^3$. We estimate the contributions
of the five noise sources discussed here as follows: 
$\tilde{\sigma}_1  \approx 5\mu K$ and 
$\tilde{\sigma}_J \approx 
20 \mu K$ for $J = 2-5$. The filtered $\tilde{\sigma}$ of intracluster
gas internal flow \cite{Nagai03}  is $\la 10 \mu K$, due to its random 
nature.  Lack of knowledge on its spatial distribution, we neglect
this term.  For SPT, the  errors caused by the cosmic 
variance and shot noise  ($\sim 10\%$) dominate over all other errors. 
The systematic errors are virtually always sub-dominant.   In this sense,
our method has the special advantage  to measure the cluster peculiar
velocity. For a 
future all sky  survey, total error can be reduced to several percent level.

\section{Conclusions}
\label{sec:conclusion}

We presented a new method to estimate the peculiar velocity power
spectrum using the KSZ effect. The signal (cluster bulk velocity
correlation) is amplified by direct auto correlation of the cluster
KSZ flux. Consequently, many systematics, such as internal flow
involved in the usual $v_p$ inversion, vanish and the majority of
remaining systematics can be simply subtracted. The correlation method
has the special advantage that statistical error dominates over systematics
at effectively all $k$ and $z$ range. 

Ideas similar to those expressed here about prospects of probing
the peculiar velocity field of clusters at high redshift through
KSZ flux correlations have been recently presented in \cite{carlos2006}.
Our papers are complementary: we focus on spatial correlations,
while \cite{carlos2006} study angular correlations.

Our calculations show that the SPT can measure $\Delta^2_v$
to $\sim 10\%$ accuracy at $k\sim 0.02$-$0.3 h/$Mpc and a future all sky
survey can improve this measurement by a factor of few.   
These results should be regarded as preliminary --
the goal of this paper was to present a new idea
and a rough estimate of the expected signal--to--noise ratio.
A comparison with more realistic numerical simulations
should be the next step.

\section{Acknowledgments}
PJZ thanks the support of  the DOE and the NASA
grant NAG 5--10842 at Fermilab at the early stage of this work and the
National Science Foundation of China grants 10673022 and 10533030. HAF 
was supported in part by the  University of Kansas GRF and by a grant
from the Research Corporation. 
RJ was supported by (to be added just before submission).
AS is  supported by 
the DOE and the NASA  grant NAG 5--10842 at Fermilab.

\end{document}